\documentclass{article}

\usepackage{arxiv}

\usepackage[utf8]{inputenc} 
\usepackage[T1]{fontenc}    
\usepackage{hyperref}       
\usepackage{url}            
\usepackage{booktabs}       
\usepackage{amsfonts}       
\usepackage{nicefrac}       
\usepackage{microtype}      
\usepackage{cleveref}       
\usepackage{lipsum}         
\usepackage{graphicx}
\usepackage{natbib}
\usepackage{doi}
\usepackage{multirow}
\usepackage{subfig}

\title{Quality and Cost Trade-offs in Passage Re-ranking Task}

\author{{Pavel Podberezko} \\
	\And
	{Vsevolod Mitskevich} \\
	\And
	{Raman Makouski} \\
	\And
	{Pavel Goncharov} \\
	\And
	{Andrei Khobnia} \\
	\And
	{Nikolay Bushkov} \\
	\And
	{Marina Chernyshevich} \\
}
\date{
{IHS Markit \\
\texttt{first.last@ihsmarkit.com}}
}

\renewcommand{\headeright}{Notebook paper}
\renewcommand{\undertitle}{Notebook paper}

\begin{document}
\maketitle

\begin{abstract}
	Deep learning models named transformers achieved state-of-the-art results in a vast majority of NLP tasks at the cost of increased computational complexity and high memory consumption. Using the transformer model in real-time inference becomes a major challenge when implemented in production, because it requires expensive computational resources. The more executions of a transformer are needed the lower the overall throughput is, and switching to the smaller encoders leads to the decrease of accuracy. Our paper is devoted to the problem of how to choose the right architecture for the ranking step of the information retrieval pipeline, so that the number of required calls of transformer encoder is minimal with the maximum achievable quality of ranking. We investigated several late-interaction models such as Colbert and Poly-encoder architectures along with their modifications. Also, we took care of the memory footprint of the search index and tried to apply the learning-to-hash method to binarize the output vectors from the transformer encoders. The results of the evaluation are provided using TREC 2019-2021 and MS Marco dev datasets.
\end{abstract}


\section{Introduction}
Transformer based models achieved state-of-the-art results on a diverse set of tasks across different domains but most of the architectures have quadratic memory and computational complexity \citep{devlin2019bert}. Some recent studies introduced models with the reduction of original complexity \citep{tay2020efficient} either in terms of memory or computational type, but nevertheless, they are slow too in comparison with the other steps of the information retrieval pipeline considered in this article. Thus, often the execution of inference of a transformer model dominates the total runtime of a prediction pipeline and it becomes a real challenge when millions of documents should be processed using the transformer as a backbone. 

Classical information retrieval pipeline includes several phases – building a searching index; vectorization of an incoming query; retrieving a top of nearest neighbors, e.g. examples from the index which are the most similar to the input query; re-ranking of the extracted top in order to provide the user with results sorted by their relevancy to query. Building a search index phase could be done offline, but vectorization of the incoming query, extraction of the top of relevant examples, and re-ranking must be executed in real-time. Re-ranking becomes the most time-consuming step if the transformer models are used for that. 

Our work addresses the problem of choosing the right architecture of the model for re-ranking in order to increase throughput without losing much of the ranking quality. By architecture we mean not the encoder model itself, but the way of obtaining the final ranking score. We investigate the application of several late-interaction models (also called split-encoders) like Colbert \citep{khattab2020colbert} and poly-encoder \citep{humeau2020polyencoders} to the task of passages re-ranking. We experiment with the type of rankers on the top of transformer backbones, the size of output vectors, and the data type for weights. Besides, we provide experiments with binarization of the output vectors from encoder models using the learning-to-hash method \citep{yamada2021efficient} to save the disk space when storing the pre-computed index for re-ranking. 
The results and their analysis are provided in the last section of the work.

\section{Related Works}
\label{sec:related_work}

Deep transformer models pretrained as language models, for example BERT \citep{devlin2019bert} or RoBERTa \citep{liu2019roberta},
have proven highly effective in a diverse set of classification and sequence labeling tasks in natural language processing. Nogueira and Cho \citep{nogueira2020passage} \citep{nogueira2019multistage} demonstrated effectiveness of deep transformer models in ranking tasks. These types of models are usually referred to as \textbf{cross-encoders}. The main disadvantage of such models is the necessity to perform full self-attention over the query-document pair, so these models are slow for practical use. Due to this fact, it is impractical to apply inference to every document in a corpus with respect to a query, so these techniques are typically applied to re-rank a rather small list of candidates. 
 
Also, there is a broad class of models that map the input and a candidate label separately into a common feature space wherein typically a dot product or cosine is used to measure their similarity. These models are usually referred to as \textbf{Bi-encoders}. Researchers also consider the hybrid type of models that use separate encoding for query and document and different types of late interaction between encoded vectors. We refer to these models as \textbf{split-architectures}. One of the examples is poly-encoders \citep{humeau2020polyencoders}. It is an architecture with an additional learned attention mechanism that represents more global features from which to perform self-attention, resulting in performance gains over Bi-encoders and large speed gains over Cross-Encoders. But authors provided experimental results only for four downstream tasks without measurements on the most popular IR benchmarks like TREC or MS MARCO \citep{bajaj2018ms}. Another example of split-architectures is ColBERT \citep{khattab2020colbert}. It introduces a late interaction architecture that independently encodes the query and the document using BERT and then employs a cheap yet powerful interaction step that models their fine-grained similarity. By delaying and yet retaining this fine-granular interaction, ColBERT can leverage the expressiveness of deep LMs while simultaneously gaining the ability to pre-compute document representations offline, considerably speeding up query processing. Experimental results of ColBERT training for MS MARCO and TREC CAR \citep{nanni2017benchmark} datasets were provided. Another approach called PreTTR (Precomputing Transformer Term Representations), \citep{PreTTR} suggested using separate encoding for query and document and then several neural layers to merge them at query time to compute the final ranking score. Like in the approaches above, this allows one to precompute part of the document term representations at indexing time. Two benchmark results are available for this approach: WebTrack 2012 and Robust 2004.

Another direction to find quality and cost trade-offs for ranking models is vector size optimization and reduction. One of the promising approaches is a vector binarization \citep{yamada2021efficient}. It is a memory-efficient neural retrieval model that integrates a learning-to-hash technique into the state-of-the-art Dense Passage Retriever. It reduces the memory cost from 65GB to 2GB without a loss of accuracy on two standard open-domain question answering benchmarks: Natural Questions and TriviaQA.

\section{Methodology}
\label{sec:methodology}

We trained our models on MS Marco passage ranking dataset using triples.train.small
from the official website\footnote{https://microsoft.github.io/msmarco/}. The data consists of around 40M triples in the format (query,
positive passage, negative passage), where positives for each query were taken from
sparsely annotated index of 8.8M passages, while negatives are sampled from passages
retrieved by bm25 model. It is well known that such approach leads to the large amount of
false negative examples, so besides training on triples and considering first passage
as positive and the last one as negative, we also tried distillation based training where
labels for train examples were generated by other model. For this work we used labels
generated by \cite{hofstaetter2020_crossarchitecture_kd} and available at authors' github\footnote{https://github.com/sebastian-hofstaetter/neural-ranking-kd}.
These labels are logits from the ensemble of three cross-encoder models: bert-base, bert-large and albert-large.

\subsection{ColBert}\label{sec:met_colb}
One of the architectures chosen for this work is ColBert, which is the late interaction model
with separate vectorization of a query and a passage.

Confirming results from the \cite{khattab2020colbert} where ColBert was introduced, we found `[MASK]`
augmentation of query to be helpful in terms of final ranking quality across all our experiments.
We use fixed maximum length of a query (32 tokens) and pad any shorter query to this length
with `[MASK]` token, while truncating any longer one.

ColBert uses MaxSim ranker as a scoring function, which can use L2 distance, dot product
or cosine as a similarity metric between two vectors. In our experiments best results were
achieved with L2 distance, so further in this paper ColBert is presented only with L2 ranking function. 

Also, comparing to the original implementation in \cite{khattab2020colbert}, we found that additional
LayerNorm after hidden states dimension reduction layer (for example, from 768 of bert-base to
128 as in original ColBert implementation) generally improves quality throughout most of
the experiments, so any further results with ColBert in this paper are presented with this layer included.

Another degree of freedom in Colbert model is the size of output vectors for a query and
a passage. In this work we carried out several experiments with sizes of 128, 64, 32 to study
possible trade-offs between ranking quality, vectors storage cost and response latency.

Finally, to achieve best possible quality in terms of ranking metrics, while maintaining
low response latency, we trained ColBert model with a range of transformer encoders as
its backbone. Among those is T5, which has different tokenizer compared to Bert and Bert-like
models (Electra, Albert, etc.), and this leads to a few subtleties in ColT5 implementation.
First, T5 tokenizer does not have `[MASK]` token, so instead of it for query augmentation we
tried several options and found `<unk>` token to be the best replacement. Second, `</s>`
token was inserted instead of usual `[SEP]` and no replacement for `[CLS]` was used.

\subsection{Applying learning-to-hash to Colbert}
One of the experiments we have done is the application of learning-to-hash approach to our ColBert models.
The objective of this research is to reduce index memory size, and evaluate trade-off between
size and ranking metrics. Our implementation mostly follows the approach introduced in \cite{yamada2021efficient}.
We have modified two-stage scheme from original paper into one-stage, which means binarization
of re-ranking stage only.

\subsection{Poly-encoder}
Also we explore Poly-Encoder model. We trained the classic Poly-Encoder described in the \cite{humeau2020polyencoders}.
In the classical model, we found that the number of codes does not play a big role with our requests, because the requests had a limited length (about 10 tokens), so we settled on 8 codes. As an initial checkpoint in Poly-Encoder training Castorini TCTColbert \citep{lin2020distilling} was used.

\section{Experimental Setup}
All models were trained on ~40M triples of MS Marco triples.train.small for one epoch. 
For evaluation, the following datasets for passage reranking task were used:
\begin{itemize}
	\item TREC 2019 DL and TREC 2020 DL: test sets with 43 and 54 densely judged queries correspondingly with candidates scores varying between 0 and 3.
	\item MS Marco dev: 6980 sparsely judged queries with typically one relevant passage per query.
	\item TREC 2021 DL: 53 densely judged multiple graded relevance labels.
	Three runs (ihsm\_colbert64, ihsm\_bicolbert and ihsm\_poly8q) were submitted
	 to top-100 passage reranking task of TREC 2021.
\end{itemize}
Official lists of top candidates selected with bm25 algorithm and provided by datasets authors were used.

For all runs we set the maximum length of a query to be 32 tokens and for the passages 192 tokens. As optimizer, AdamW was used, learning rate was set to 2e-5 for the weights of pretrained transformer and to 5e-5 for new weights (such as linear compression layer in ColBert model). Typically for most runs batch size for base models was 32 and the number of gradient accumulation steps was set to 4.
For all our experiments we used pytorch and transformers libraries. Training was performed in automatic mixed precision setting.

\section{Results and Analysis}
Our main results are presented in Table~\ref{tab:table}. It shows ranking metrics on four evaluation datasets,
as well as latency per query for reranking task. The latency is measured on single RTX 2080 TI GPU and
Intel Xeon E5-2680 (restricting pytorch to use 8 threads)
and shown in two components: query vectorization and candidates reranking when all vectors are already loaded in RAM.

ColT5-base on the first row is the experiment with T5Encoder model as a backbone for ColBert model.
It was trained on MS Marco triples using cross-entropy loss and gave the best result among all tested
transformer encoders in the setting without distillation.
All the rest rows in the table show results for experiments with distillation, when margin-mse loss
was used on labels described in section~\ref{sec:met_colb}.

\begin{table}[b!]
	\caption{Re-ranking quality achieved with different models on four datasets.
	Latency both for CPU and GPU is shown as well.
	The last column presents throughput measurements on GPU.}
	\centering
	\renewcommand{\arraystretch}{1.6}
	\resizebox{\textwidth}{!}{\begin{tabular}{llrrrrrrrrrr}
		\toprule
		\multirow{3}{*}{Row} & \multirow{3}{*}{Model} & \multirow{3}{1cm}{Vector size} & \multirow{3}{1.5cm}{\raggedright TREC 2019 DL ndcg10} & \multirow{3}{1.5cm}{\raggedright TREC 2020 DL ndcg10} & \multirow{3}{1.5cm}{\raggedright TREC 2021 DL ndcg10} & \multirow{3}{1.5cm}{\raggedright MS Marco dev mrr10} & \multicolumn{4}{p{5cm}}{\centering Latency, ms/query (1000 passages)} & \multirow{3}{1.5cm}{\raggedright Index vectorization, ms/passage} \\
		\cmidrule(r){8-11}
		 & & & & & & & \multicolumn{2}{c}{GPU} & \multicolumn{2}{c}{CPU} & \\
		 \cmidrule(r){8-11}
		 & & & &  & & & Q vec & Rank & Q vec & Rank & \\
		\midrule \midrule
		1 & ColT5-base & N$\times$128 & 0.6903 &	0.7101	& - & 0.3449 &	15 & 30 & 45 & 400 & 2.5 \\
		2 &	ColBert-base & N$\times$128 & 0.7253 & 0.7442 & - & 0.3743 & 15 & 30 & 45 & 400 & 2.5 \\
		\bf 3 & \multicolumn{1}{p{2.5cm}}{\raggedright \bf ColBert-base (ihsm\_colbert64)} & \bf N$\times$64 & \bf 0.7324 & \bf 0.7415 & \bf 0.6453 & \bf 0.371 & \bf 15 & \bf 15 & \bf 45 & \bf 200 & \bf 2.5 \\
		4 &	ColBert-base & N$\times$32 &	0.7408 & 0.723 & - & 0.3664 & 15 & 7.5 & 45 & 100 & 2.5 \\
		5 &	ColBert-large & N$\times$128 & 0.7347 & 0.7293 & - & 0.3715 & 35 & 30 & 120 & 400 & 7.5 \\
		6 &	\multicolumn{1}{p{2.5cm}}{\raggedright ColEctra-small-generator} & N$\times$64 &	0.7223 & 0.7169 & -	& 0.358 & 15 & 15 & 20 & 200 & 1.1 \\
		7 &	\multicolumn{1}{p{2.5cm}}{\raggedright ColEctra-large-generator} & N$\times$64 &	0.7298 & 0.7247 & - & 0.3657 & 30 &	15 & 40 & 200 & 1.8 \\
		8 &	ColMiniLM-L4 & N$\times$64 & 0.7092 & 0.7258 & - & 0.3561 & 5 & 15 & 12 & 200 & 1.1 \\
		\bf 9 & \multicolumn{1}{p{2.5cm}}{\raggedright \bf BinaryColBert-base (ihsm\_bicolbert)} & \bf N$\times$256 & \bf 0.7304 & \bf 0.7303 & \bf 0.6393 & \bf 0.3553 & \bf 15 & \bf 60 & \bf 45 & \bf 800 & \bf 2.9 \\
		\bf 10 & \multicolumn{1}{p{2.5cm}}{\raggedright \bf Polyencoder bert-base (ihsm\_poly8q)} & \bf 1$\times$768 & \bf 0.7223 & \bf 0.7205 & \bf 0.6342 & \bf 0.3315 & \bf 15 & \bf 3 & \bf 55 & \bf 5 & \bf 2.5 \\
		\bottomrule
	\end{tabular}}
	\label{tab:table}
\end{table}

Among presented experiments with distillation are ColBert runs with backbones of Bert-base,
Bert-large, Electra-generator-small, Electra-generator-large and MiniLM-L4. Besides them,
also T5Encoder and Electra-discriminator were tested, but they gave quite poor results.

For the case of T5Encoder, which shows great effectiveness without distillation, we can argue
that the differences in the range of output scores between T5 and Bert might be the cause of
failure, because distillation labels were mainly produced by Bert model.

\begin{figure}[b!]
    \centering
    \subfloat[]{\includegraphics[width=0.5\textwidth]{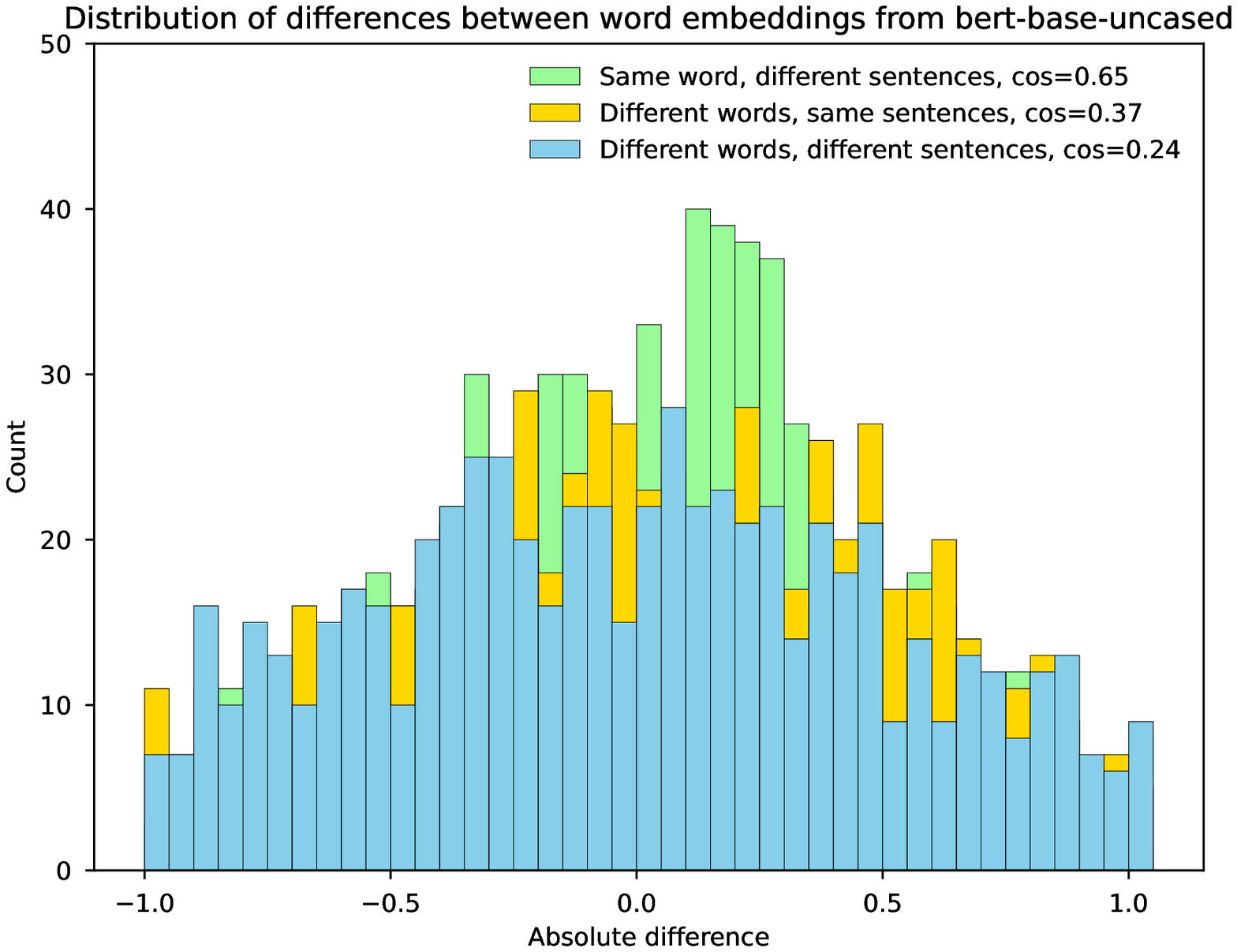}} 
    \subfloat[]{\includegraphics[width=0.5\textwidth]{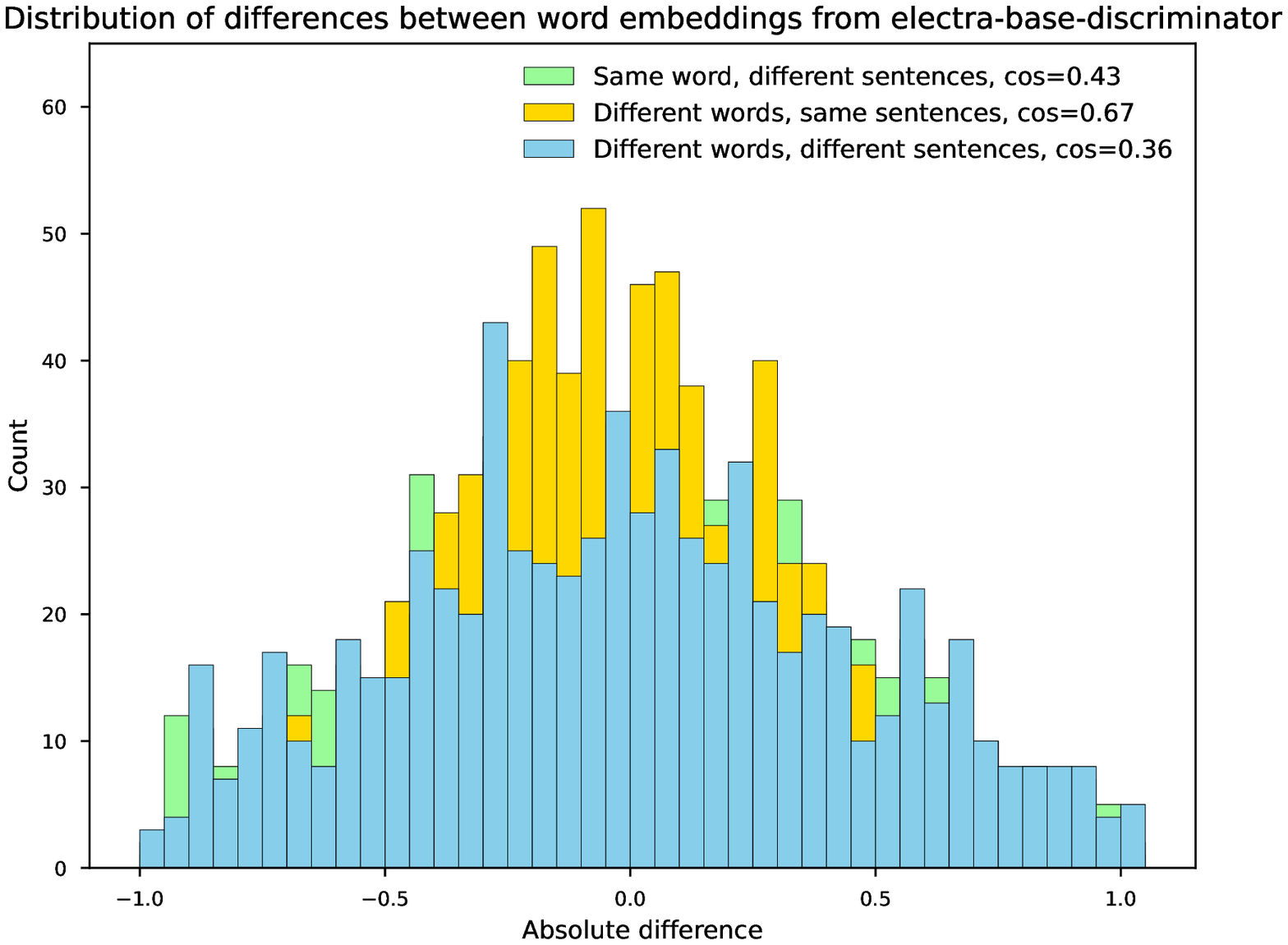}} \\
    \subfloat[]{\includegraphics[width=0.5\textwidth]{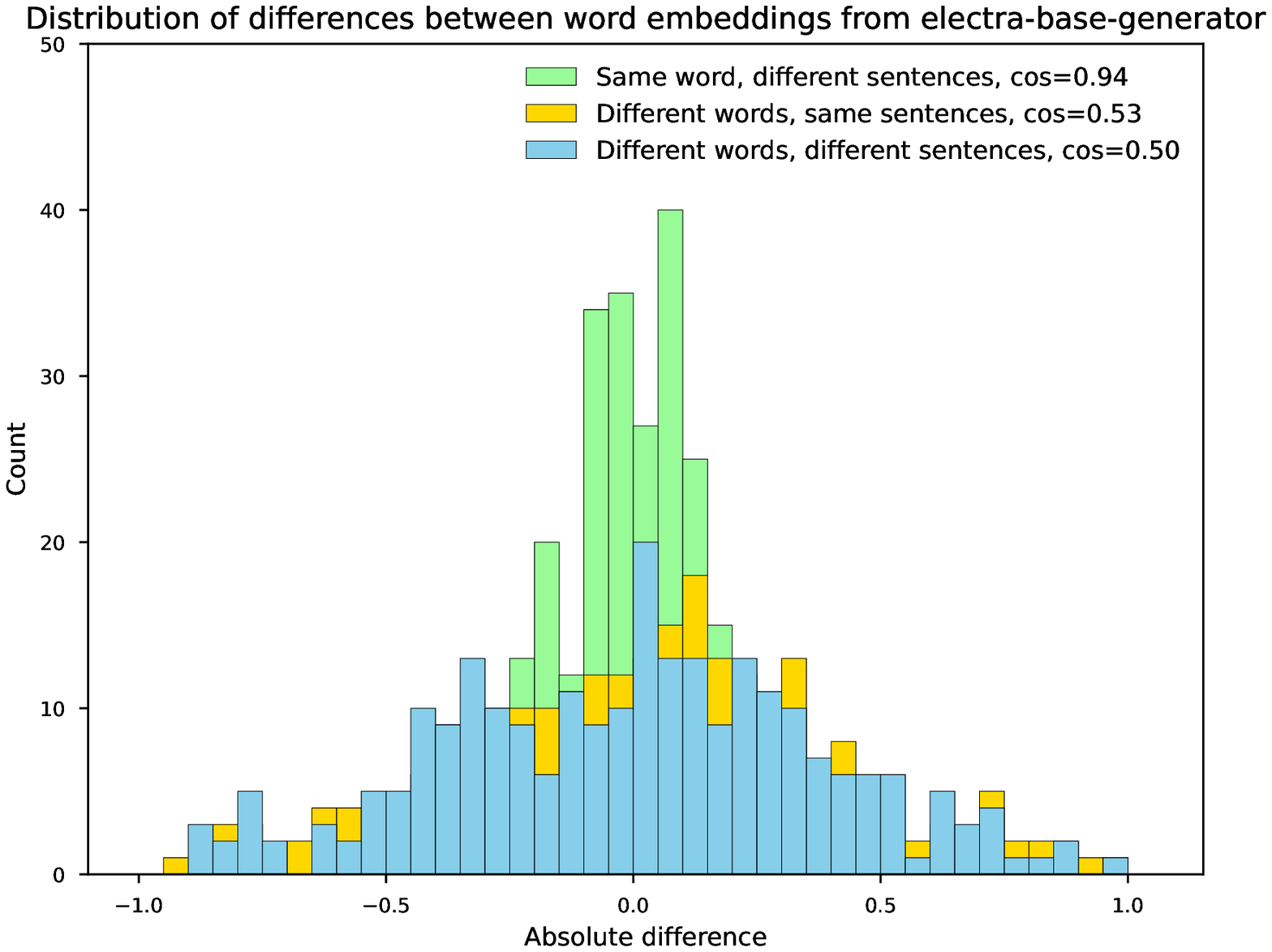}}
    \subfloat[]{\includegraphics[width=0.5\textwidth]{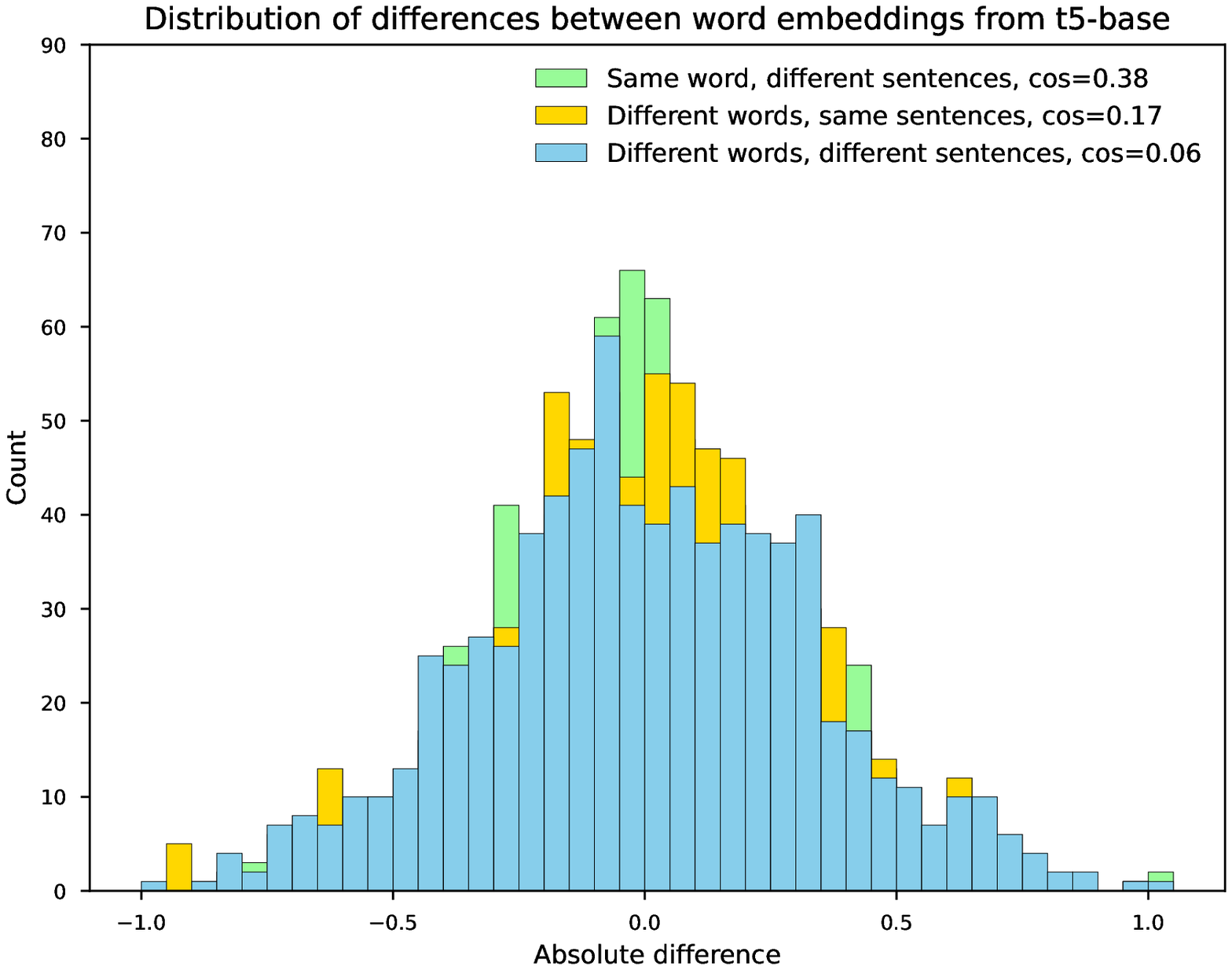}}
	\caption{Distributions of differences between word vectors from various transformers.
	Differences are calculated as vec1-vec2, where vec1 and vec2 are vectors of either different tokens
	or same token from different sentences and are taken from the last
	layer of transformer encoder. a -- bert-base-uncased, b -- electra-base-discriminator,
	c -- electra-base-generator, d -- t5-base.}
	\label{fig:dist_plot}
\end{figure}

As for Electra-discriminator model, it shows extremely low results both with and without distillation
and seems to be unsuitable choice for ColBert backbone. To understand the reason behind such behavior,
we analyzed output vectors from different transformers.
We observed that output embeddings of Electra-discriminator model are much more context dependent
than vectors from most other transformers as shown in Figure~\ref{fig:dist_plot}. The analysis was conducted as follows:
two example sentences were taken, which share at least one word in common and the meaning of that word
is the same in both sentences. Then the values of the vector of difference between embeddings of this
word from the first and the second sentence are calculated (plotted as green distribution in Figure~\ref{fig:dist_plot}).

After that, any other word, which has clearly different from meaning from the previously selected word,
is taken from the first sentence and its embedding is compared in the same way with the embedding of
the first word produced both in the context of the first sentence (yellow histogram – different words,
same sentences) and in the context of the second sentence (blue histogram – different words, different
sentences). Namely, plotted here are the results of calculation for the following example sentences
from Wikipedia: 1) “Discoveries in organometallic chemistry have led to important insights into
chemical bonding.”, 2) “The 18-electron rule is the equivalent of the octet rule in main group
chemistry”. The word, which is compared between sentences is “chemistry”, the word with different
meaning is “have”. Plots a), c), and d), which represent bert-base-uncased, electra-base-generator
and t5-base correspondingly, show that for all these models the word “chemistry” is closer to itself
from different contexts, than to the word “have”, while plot b) for electra-base-discriminator shows
that this model considers “have” to be closer to “chemistry”, when this word are used in the same
sentence, than “chemistry” from a different sentence.

The observed peculiarity of Electra-discriminator’s output embeddings might be the
explanation of its poor results, because ColBert’s ranker heavily relies on token-to-token
similarity from different contexts.

Experiments presented in the Table~\ref{tab:table} on rows from 2 to 4 deal with the size of the output vectors
and mainly investigate the trade-off between effectiveness and the storage cost. ColBert with the
largest output dimension of 128 is indeed seems to be the best in terms of ranking quality,
but the model with 64-dimensional vectors is not that far behind, with this gap being slightly
smaller than the one between 64- and 32-dimensional models. So overall vectors size of 64 looks like
the optimal value to balance between quality and costs and this variant ColBert-base was chosen
as a submission to TREC 2021 DL.

Experiment on the 5th row presents results for the bert-large model as a backbone for ColBert.
As one can see, it does not show any advantage over the base model in terms of ranking quality,
but performs vectorization of query significantly slower. Lack of quality improvement might be
caused by suboptimal selection of hyperparameters for training as larger models tend to be more
sensitive to the variations in training conditions.

The following lines from 6 to 8 show results for the models with fewer parameters than bert-base.
ColEctra-large-generator (which is ColBert with electra-large-generator as a backbone encoder)
with 64-dimensional vectors is quite close in ranking quality to the result of ColBert with
32-dimensional vectors, while having 51M parameters compared to 110M in bert-base-uncased.
2 times larger output embeddings mean 2 times larger storage costs and 2 times slower online
ranking operation. Also, ColEctra-large-generator is 2 times slower than ColBert-base in query
vectorization on GPU due to its encoder having twice as many layers as bert-base-uncased.
Nevertheless ColEctra-large-generator has two advantages over ColBert-base: firstly, it is roughly
40\% faster in offline index vectorization on GPU (measured on RTX 2080 TI) and secondly,
it is 15\% faster in online query vectorization when it has to be performed on CPU (measured on Intel
Xeon E5-2680), both advantages emerging from the smaller number of parameters.
ColEctra-small-generator with only 13.5M parameters pushes speed gains even further: both offline
index vectorization on GPU and online query vectorization on CPU are roughly 2.2 times faster
compared to ColBert-base. Though the drop in ranking quality is much more pronounced in this case
when compared to the best results of ColBert-base, the result is still decent and better than we
the best we could achieve with bert-base sized models in training setting without distillation.
The effectiveness of ColMiniLM-L4, based on MiniLM-L4 model with only 4 layers and 19M parameters,
is very similar to that of ColEctra-small-generator. The speed of offline index vectorization
on GPU is also roughly the same. But due to fewer layers it is faster in online query vectorization
both on GPU and CPU than any other model presented in this work. Comparing to ColBert-base,
the gain is of factor of 3 for GPU and up to 5 for CPU vectorization.

The 9th row of the Table~\ref{tab:table} represents our experiment with application of
learning-to-hash technique to ColBert. Binarization reduces the size of output embeddings by the factor of 32
compared to the output of standard ColBert (which produces vectors in float32) and dramatically reduces storage costs. But in our experiments binarized
ColBert showed quite significant drop in ranking quality. In order to mitigate this, we increased the number of
elements in vectors from 64 to 256, which helps in reaching competitive quality. When comparing to ColBert-base-64, the presented result has 1.5\% lower mrr10 score on MS Marco dev dataset and its re-ranking stage is 4 times slower due to larger vectors, but it reduces storage costs by the factor of 8. Among our submits to TREC 2021 DL, ihsm\_bicolbert gives the lowest score in terms of ndcg10, but margins are quite small and the gap to ihsm\_colbert64 is only around 1\%.

The last line of the Table~\ref{tab:table} is the result of poly-encoder model with 8 attention codes. As one can see, two main advantages of this model are fast ranking function and low storage costs both due to keeping only one vector per index passage. The result on MS Marco dev dataset shows that such benefits come at the cost of quality drop below the level of ColT5 model trained without distillation. Nevertheless, according to the ranking metrics of TREC DL 2019-2021 poly-encoder has competitive results, being only about 2\% below ColBert. Overall, poly-encoder can be considered as a reasonable choice for some search tasks due to its practical benefits.

\section{Conclusion}
We trained and evaluated a bunch of re-ranking models on popular passage re-ranking datasets and studied the quality-cost trade-offs for various design choices. The best quality can be achieved with ColBert architecture and bert-base model as its backbone, setting output vector size to 64: either increasing this size or choosing larger bert variant does not lead to significant improvement in ranking metrics. In order to reduce storage costs, one can reduce vector size by a factor of 2 with rather small loss of ranking quality. On the other hand, to increase inference speed (on CPU for online vectorization and on GPU for offline) one can change encoder model to electra-generator-larger with similar trade-off in quality. We showed that large storage costs reduction with reasonable ranking metrics can be achieved with binarization of ColBert output vectors or by opting for Poly-encoder model. Also we confirmed the result of \cite{hofstaetter2020_crossarchitecture_kd} that cross-architecure knowledge distillation can dramatically improve the end quality of smaller models. Large latency reduction can be achieved with such models as electra-generator-small or MiniLM-L4 serving as a ColBert's backbone, while ranking metrics will be maintained at the level of larger models trained without knowledge distillation.

\bibliographystyle{unsrtnat}
\bibliography{references}  






\end{document}